\documentclass[letterpaper]{article} 
\usepackage{aaai2026}  
\usepackage{times}  
\usepackage{helvet}  
\usepackage{courier}  
\usepackage[hyphens]{url}  
\usepackage{graphicx} 
\usepackage{natbib}  
\usepackage{caption} 
\usepackage{placeins} 
\usepackage{algorithm}
\usepackage{algorithmic}

\usepackage{newfloat}
\usepackage{listings}
\DeclareCaptionStyle{ruled}{labelfont=normalfont,labelsep=colon,strut=off} 
\floatstyle{ruled}
\newfloat{listing}{tb}{lst}{}
\floatname{listing}{Listing}

\title{Can LLMs Identify Tax Abuse?}
\author {
    Andrew Blair-Stanek\textsuperscript{\rm 1, 2},
    Nils Holzenberger\textsuperscript{\rm 3},
    Benjamin Van Durme\textsuperscript{\rm 2}
}
\affiliations {
    
    \textsuperscript{\rm 1}University of Maryland Carey School of Law\\
    \textsuperscript{\rm 2}Johns Hopkins University\\
    \textsuperscript{\rm 3}T\'el\'ecom Paris - Institut Polytechnique de Paris\\
     ablair-stanek@law.umaryland.edu, nils.holzenberger@telecom-paris.fr, vandurme@jhu.edu
}

\begin{document}

\maketitle

\begin{abstract}
We investigate whether large language models can discover and analyze U.S. tax-minimization strategies. This real-world domain challenges even seasoned human experts, and progress can reduce tax revenue lost from well-advised, wealthy taxpayers. We evaluate the most advanced LLMs on their ability to (1) interpret and verify tax strategies, (2) fill in gaps in partially specified strategies, and (3) generate complete, end-to-end strategies from scratch. This domain should be of particular interest to the LLM reasoning community: unlike synthetic challenge problems or scientific reasoning tasks, U.S. tax law involves navigating hundreds of thousands of pages of statutes, case law, and administrative guidance, all updated regularly. Notably, an LLM identified an apparently novel tax strategy, highlighting these models' potential to revolutionize tax agencies' fight against tax abuse.
\end{abstract}

\begin{links}
    \link{Code \& Data}{https://github.com/BlairStanek/LLM-Tax-Abuse/}
\end{links}

\section{Introduction}
\label{intro}

\begin{figure}[ht]
\centering
\includegraphics[width=2.8in]{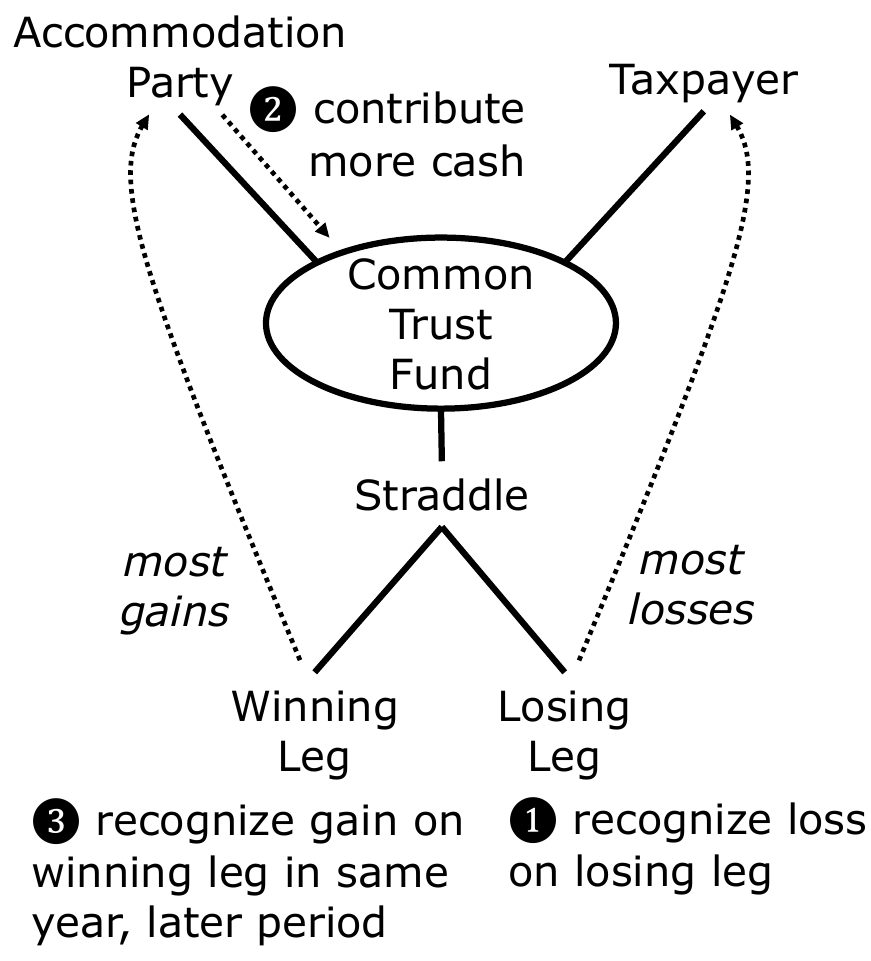}

\caption{Diagram of a novel tax strategy generated by o1-preview during our experiments. Creating such strategies normally requires extensive effort from expensive, specialized domain experts. While we find no LLMs consistently generating workable strategies, to our knowledge this is the first time an LLM has invented a tax strategy.}
\label{fig_novel}
\end{figure}

Tax-minimization schemes cost the U.S. government over \$600 billion annually \citep{IRStaxgap}.  Other governments around the world also lose substantial tax revenue to such schemes \citep{OECD_beps}.  This problem is uniquely susceptible to being solved by artificial intelligence, since it involves huge amounts of data and millions of words of tax-law authorities, beyond the capacity of any one individual to master.  But to fulfill this promise, AI must be able to reason over the text of tax-law authorities, understand how tax strategies work, and generate possible tax strategies based on existing tax-law authorities.  We benchmark several leading LLMs' abilities to do this; we find them generally mediocre but with occasional flashes of brilliance.

We introduce a new domain-expert crafted dataset called Shelter Check, derived from public sources that describe schemes used in the past by taxpayers to minimize taxes.  The dataset includes the text of the tax-law authorities, like tax code sections or court decisions, that enabled the scheme. It also includes: the background facts necessary for a taxpayer to use the scheme; the goals of the scheme (e.g. creating an artificial tax deduction); the steps of the scheme itself; and a legal analysis of why the scheme arguably meets the goals, given the tax-law authorities.  We also include an adversarial step that, if used by a taxpayer, would make the strategy unquestionably fail to meet at least one of the goals.

We use Shelter Check to thoroughly test the abilities of three leading reasoning models: OpenAI's o3, Anthropic's Claude-4, and Google's Gemini-2.5.  As a baseline, we also run the tests against a non-reasoning, open-weight model, the 70B parameter version of Meta's Llama-3.3-Instruct.  

As an initial step, we test whether the LLMs can validate each step of our legal analysis of each strategy, evaluating whether it is correct or merely viable.  More substantively, we test whether the LLMs can analyze whether the stated goals are met by the strategy steps, given the background facts and textual authorities.  To test the LLMs' abilities to identify tax strategies that do \textbf{not} work, we repeat this experiment, using the adversarial step that should prevent the strategy from working.

We then evaluate the ability of the LLMs to generate tax strategies.  We try a ‘‘step-cloze" task where we ask the LLMs to fill in a single blanked-out step in our known tax strategies.  We also asked LLMs to generate entire tax strategies from scratch, prompting them with only the background, goals, and authorities behind known past strategies.  A domain expert graded these model-generated strategies; we also had other models grade these strategies and compared the models' grades against the domain expert's grades.

Notably, o1-preview generated what appears to be an entirely novel working tax-minimization strategy, which is depicted in Figure~\ref{fig_novel}, discussed in section~\ref{sec_from_scratch}, and posted on GitHub.  This result suggests LLMs might have revolutionary potential for helping governments find and thus shut down such strategies.

\section{Background}
\label{background}

Automated legal reasoning has a history dating back to expert systems. Legal rules can be formalized into Prolog-style expert systems that compute answers to straightforward legal questions \citep{mccarty77reflections,sergot86british}. The formalization makes explicit some amount of common sense knowledge and implicit assumptions, so that this step may require the involvement of a legal expert \citep{merigoux21catala}. This rule-based reasoning can be naturally extended to case-based reasoning, provided the applicable rules (analogy, proportionality, etc.) are formalized too \citep{gardner83design,bruninghaus03predicting,gray24using}.  \citet{cao_rr} focus on detecting tax evasion in raw data. 

In principle, a perfect translation of all tax authorities from natural language into computer-interpretable code would make the search for tax strategies a matter of 
running a closed-world simulation. One example of such a simulation is the work of \citet{hemberg2015tax} and \citet{hemberg16detecting}. They model the relationship between tax evasion and auditing policies, and how each dynamically adapts to the other's behavior. But the scalability of such efforts is unclear, given the effort required to translate just a few sections of tax law into computer-interpretable form.

\citet{fratric2025aiexposetaxloopholes} let virtual agents explore the full scope of possibilities allowed by a formal model of the law, constructed through a costly process of translating law to code. The economic profit associated with a specific sequence of agent actions is then used as an indicator for potentially illegal tax strategies.

But legal authorities are written in natural language, potentially making it preferable to engage directly with the legal text.  
Several datasets have been built around tasks involving legal text. LexGLUE \citep{chalkidis-lexglue} groups multiple English-language legal NLP datasets. All tasks are classification tasks, except for one multiple-choice QA task. Analogous datasets exist for other languages and jurisdictions, typically framed as classification or retrieval tasks: judgment prediction for Indian courts \citep{malik-ildc} and Swiss courts \citep{niklaus-swiss}, multiple choice questions from the National Judicial Examination of China \citep{zhong-jec-qa}, and retrieval of statutes and cases for Japanese law \citep{goebel-overview}, \emph{inter alia}. Closer to our dataset, LegalBench \citep{guha23legalbench} is a collection of legal tasks, some from existing datasets, some contributed by legal experts. The tasks expect a wide variety of output formats, and are meant to test LLM abilities. In particular, LegalBench contains two tasks that explicitly involve tax: the statutory reasoning assessment (SARA); and predicting outcomes in cases from the Tax Court of Canada. Both are framed as classification tasks. While state-of-the-art LLMs reach high performance on the self-contained benchmarks cited above, \citet{nay2024large} have further documented LLM abilities closer to the practice of tax law, reporting mixed results.
Similarly, \citet{zhou2024rulearena} benchmark LLMs' abilities to apply several rulesets, one of which is a curated subset of the basic U.S. tax regulations; they also find mixed performance.
\citet{hu_jh} further show that LLMs may be overly sensitive to language patterns as opposed to legal reasoning, and \citet{chen_rethinking} even suggest LLMs may be incapable of legal reasoning because they lack causal reasoning abilities. 

We previously discussed the theoretical possibility of using AI to proactively find tax strategies in a collection of tax authorities \citep{sheltercheck_2022}. This task is related to other AI challenges involving reasoning with large-scale data and combinatorial search spaces. The first chess engines prominently involved rule-based heuristics to explore an otherwise huge search space~\citep{campbell-deep}. A similar search problem occurs in attempts to automate scientific discovery: papers' conclusions can be combined to identify potentially useful future research directions \citep{evans10machine, si2025ideationexecutiongapexecutionoutcomes}. 
Another example is predicting how proteins fold~\citep{abramson24accurate} and designing bioactive molecules~\citep{li_gxvaes}. 

Here, we take a similar approach, letting an LLM predict a solution, without any explicit search procedure.
Closer to our paper is the automation of public policy design~\cite{xing_black}. 
Reliably designing tax strategies may require strong planning abilities, which current LLMs may lack \citep{aghzal-survey}.

\section{The Shelter Check Dataset}

The Shelter Check dataset currently contains 36 known U.S. tax-minimization strategies that depend upon applying textual authorities.  Why only 36?  Tax-law domain experts (i.e. U.S. tax lawyers) are specialized and in demand from corporations and wealthy individuals.  Many have an hourly billing rate over \$1000.  Each of the 36 strategies required a minimum of 9 domain expert hours to construct and verify. 

Not all tax strategies rely on textual authorities.  Some depend instead on uncertainty in asset valuation \citep{SoledThomas2023}. Such strategies are rooted in mathematical uncertainty, not legal language, and are not in our dataset.

Most of our dataset -- 29 of the 36 -- comes from the public registry of ‘‘listed transactions" maintained by the U.S.'s tax authority, the Internal Revenue Service \citep{listed_transactions}.  We excluded those listed transactions that depend on uncertainty in asset valuation.  Four of the 36 came from high-profile public decisions of U.S. courts where the IRS challenged text-based tax strategies.  Two of the 36 are highly-publicized flaws in the language of the massive tax bill passed in 2017.  The final one of the 36 was detailed in an IRS news release.

These public sources never describe the tax strategies in sufficient detail to actually implement them. Understandably, the IRS does not want to give roadmaps to unscrupulous taxpayers. Translating the public sources into actual step-by-step strategies took much domain expert time.

Each strategy is a text file with a series of sections, described below.

\subsection{Authorities}

The first section of the text file for each strategy consists of the actual text of the tax-law authorities on which the tax strategy is built.  Many of these textual authorities are sections or portions of sections of the U.S. tax code, which is the compilation of tax statutes passed by the U.S. Congress.  Other types of textual authorities included are Treasury Regulations and revenue rulings promulgated by the IRS, tax treaties the U.S. has with other countries, and case law published by courts deciding tax-law matters.  Many of these authorities are quite long, on the order of tens or even hundreds of pages. So they have been edited down to the portions necessary for the tax-minimization strategy, with ellipses (i.e. ...) showing where portions have been edited out.

\subsection{Background}

These are the background facts necessary for a taxpayer to be in a position to use the strategy.  The number of background facts ranges between one and seven, with a mean of 2.5.  An example of a background fact, from strategy 1 is ‘‘FP is a foreign person who owns an Asset with a built-in loss, with a \$100 basis and fair market value of \$20." Without such a foreign person, the entire strategy cannot work.

\subsection{Goals}

These are the goal or goals of the strategy.  The number of goals ranges between one and four, with an average of 1.4.  One of the goals always involves minimizing the present value of taxes owed by the taxpayer.  When there are other goals, sometimes they involve the taxpayer's economic position, such as receiving cash.  Sometimes other goals involve proper incentives for another party to the strategy.  For example, strategy 1 has two goals: ‘‘Have U.S. Taxpayer T recognize a loss of \$80" and ‘‘Leave FP with property rights with the same value as the fair market value of the Asset."

\subsection{Strategy}

These are the concrete steps taken to use the authorities to reach the goal or goals.  The steps are granular enough that a human tax lawyer would know how to implement them.  The number of steps ranges between two and nine, with a mean of 4.3 steps per strategy. Across all 36 strategies, there are a total of 153 strategy steps.  In strategy 1, as an example, there are five steps, with step 1 being  ‘‘FP contributes the Asset to a Trust" and step 4 being ‘‘T is given a power exercisable solely by himself to vest the corpus or the income of Subtrust in himself."

\subsection{Analysis}

This provides legal analysis about how the authorities apply to the background and strategy steps to result in arguably meeting the goal or goals.  Unlike in a traditional legal memorandum, the analysis is broken into numbered steps to facilitate iterative verification by LLMs.  The number of analysis steps ranges between two and eleven, with a total of 193 across all 36 strategies.  As an example, strategy 1's first analysis step is ‘‘When it receives it, Trust has basis \$100 in the Asset under \S 1015(b)", referring to the Trust mentioned in the strategy and tax code section 1015(b), which is in the authorities section.

\subsection{Adversarial Step}

For each of the 36 strategies, we created a single adversarial step that can be substituted to replace a step in the actual strategy steps, so that the altered strategy definitively fails to meet all the goals.  It is for testing LLMs' abilities to identify strategies that do not work.  For example, in strategy 1, the adversarial step would replace step 4 of the actual strategy with ‘‘FP is given a power exercisable solely by FP to vest the corpus or the income of Subtrust in FP".  But doing this prevents the taxpayer from reducing their taxes.

\subsection{Primary Tax Law Area}

This is a simple classification of the strategy's primary tax-law area, which can be ‘‘Income Tax" (10 strategies), ‘‘Partnership" (5), ‘‘Corporate" (4), ‘‘International" (9) and ‘‘Employee Benefits" (8).

\subsection{Strategy Type}

\begin{table*}[!ht]
  \centering

  \begin{tabular}{|l|cc|cccc|cc|}
    \hline
& \multicolumn{2}{c|}{\textbf{Analysis Verification}} & \multicolumn{4}{c|}{\textbf{Goal Verification}} & \multicolumn{2}{c|}{\textbf{Adversarial Step}} \\
         &      &
     & \multicolumn{2}{c}{\textbf{with analysis}} & \multicolumn{2}{c|}{\textbf{without analysis}} & \multicolumn{2}{c|}{\textbf{(without analysis)}} \\
     & viable & correct & viable & correct & viable & correct & viable & correct\\
    \hline
    o3 &
    193/193
  & 166/193
 
    & 36/36 
    & 29/36 
    & 35/36 
    & 26/36 
    & 12/36 
    & \phantom{0}0/36 \\

    claude-4 &
193/193 
 & 170/193 
 & 36/36 
    & 33/36 
    & 35/36 
    & 22/36 
    & 20/36 
    & \phantom{0}0/36
   \\

    gemini-2.5 &
   185/193 
 &  157/193 
    & 34/36 
    & 27/36 
      & 34/36 
      & 31/36 
      & 14/36 
      & \phantom{0}0/36 
    \\

        llama-3.3-70B &
   192/193 
 &  178/193 
    & 36/36 
    & 33/36 
      & 36/36 
      & 26/36 
      & 31/36 
      & 11/36 
    \\

    \hline
  \end{tabular}
  \caption{Results of the first three tests. Analysis verification runs for each of the 193 analysis steps. Goal verification and adversarial step goal-failure verification run for all 36 strategies. The adversarial steps are designed to make the strategy not viable, so a model that performs as well as a domain expert would find 0/36 in both rightmost columns.  
   }
  \label{tab:goal_verification}
\end{table*}

The economist \citet{stiglitz1985general} identified three high-level classifications for tax avoidance strategies:  arbitrage between taxpayers, where income is shifted from a high-tax taxpayer to a low-tax or tax-exempt taxpayer; tax-rate arbitrage, such as shifting income from a high tax rate to a lower tax rate, such as the capital gains rate; and deferral, where tax obligations are pushed to later years, reducing the present value of the tax obligations.  To these three, we add a fourth category, legal cleverness, where taxpayers cleverly use the text of legal authorities to avoid taxes entirely, without using mere deferral or arbitrage.  Stiglitz, an economist, likely did not consider legal cleverness, which is hard to model using economic models.  Of the 36 strategies, 21 involve legal cleverness, 10 involve deferral, and 5 involve arbitrage between taxpayers; none involve tax-rate arbitrage.

\section{Experiments and Results}

Our experiments are motivated by evaluating  LLMs' abilities to handle key roles in a system that automatically identifies potential tax-minimization schemes.  Since tax agencies have limited resources, only strategies meriting action should be brought to the fore.  Thus, LLMs might be prompted to act as critic models, accurately evaluating which potential strategies might work and which would not.

We systematically used our dataset to test o3, claude-4, gemini-2.5, and, as a baseline, llama-3.3-70B.  (Specifically, we used o3-2025-04-16, claude-sonnet-4-20250514, gemini-2.5-pro, and llama-3.3-instruct-70B.)

All calls were by batch API. For o3, we set the ‘reasoning effort' parameter to high. For claude-4, we set the budget for thinking tokens to 8,000 per call.  Otherwise, all parameters were left at their default. All our code is on GitHub. 

\subsection{Analysis Verification}

We test whether an LLM can understand the legal analysis steps in the Analysis section of each strategy.  The single prompt passed to the LLM consists of all the strategy's authorities, background, and strategy steps, followed by a query testing each step of the analysis.  We pose this query in two different ways.

In one variation, which we call \textbf{viable}, the query is: ‘‘Even if it is not necessarily correct, is the following analysis potentially viable, based on the information and authorities provided?"  This corresponds to the U.S. tax-law standard for a disclosed tax strategy to potentially avoid tax penalties: there being a ‘‘reasonable basis" for the strategy, based on one or more tax-law authorities.  Most taxpayers avoid strategies falling short of this standard.

In the other variation, which we call \textbf{correct}, the query is: ‘‘Is the following analysis correct based on the information and authorities provided?"  This standard is much more stringent than viability.  In practice, many tax-minimization strategies are viable (meaning they may avoid penalties) but not correct under the law (meaning they would fail if tested in court).  Taxpayers hope not to be audited and not to have the correctness of their strategy questioned.

Human tax advisors pushing a tax strategy often work iteratively with a colleague to sharpen their legal analysis to have a ‘‘reasonable basis" justifying the strategy, making it viable.  We did something similar, passing the analyses to o3, claude-4, and gemini-2.5 and addressing all legally valid concerns the models raised.  All of o3 and claude-4's concerns could be addressed, resulting in them finding all 193 analysis steps viable.  But gemini-2.5 raised some nonsensical concerns.

For example, when an analysis step referred to dollar amounts calculated ‘‘solely" under \S 951(a)(1)(A), gemini-2.5 insisted that the amount was incorrect because it did not also consider the amounts under \S 951(a)(1)(\textit{B}).  As another example, gemini-2.5 lacked the background knowledge that dividends paid to a U.S. corporation by its wholly-owned U.S. subsidiary are tax-free.  Because of such issues, gemini-2.5 found only 185 of the analysis steps viable.

In general, one would expect more analysis steps to pass the viable standard than the more stringent correct standard.  We observe this in the two leftmost columns in Table~\ref{tab:goal_verification}.

\subsection{Goal Verification (With and Without Analysis)}

We test whether LLMs can properly determine whether a tax strategy meets its goals.  The single prompt to the LLM in this task starts with the authorities, background, and strategy.  Then, the prompt either does or does not include the domain expert-written analysis.  Finally, the prompt asks either whether it is \textit{viable} that the goal is met, or whether it is \textit{correct} that the goal is met.  As a result, there are four variations on goal verification, seen in  Table~\ref{tab:goal_verification}.  When a strategy has more than one goal, we do a separate prompt for each, and report the strategy as having its goals verified if the model responds to the prompts for all goals with yes.

We would expect that LLMs would find all strategies' goals to be viable when the prompt includes all the analysis that the LLMs finds viable.  We observe that in the column under ‘‘with analysis" and ‘‘viable", where o3 and claude-4  verified 36/36.  Not surprisingly, gemini-2.5, which found only 185 of 193 analysis steps viable, also found not all strategies' goals viable.  

Since correct is a more stringent standard than viable, we would expect LLMs to verify more goals under the viable standard than under the correct standard.  We observe this for all models, when comparing the viable and correct columns.

We also test whether the models can verify goals being met without our detailed legal analysis explaining how the authorities apply to the facts.  The results are in the two columns under ‘‘without analysis" in Table~\ref{tab:goal_verification}.  Both o3 and claude-4 find fewer strategies viable or correct when the prompt lacked the analysis.  Interestingly, gemini-2.5, which is so skeptical of analysis steps, found \textit{more} strategies (31) correct when \textit{not} given the analyses, than when given them (27).  It is less skeptical of strategies when the prompt does not give the underlying analysis.

\begin{figure*}
\includegraphics[width=\linewidth]{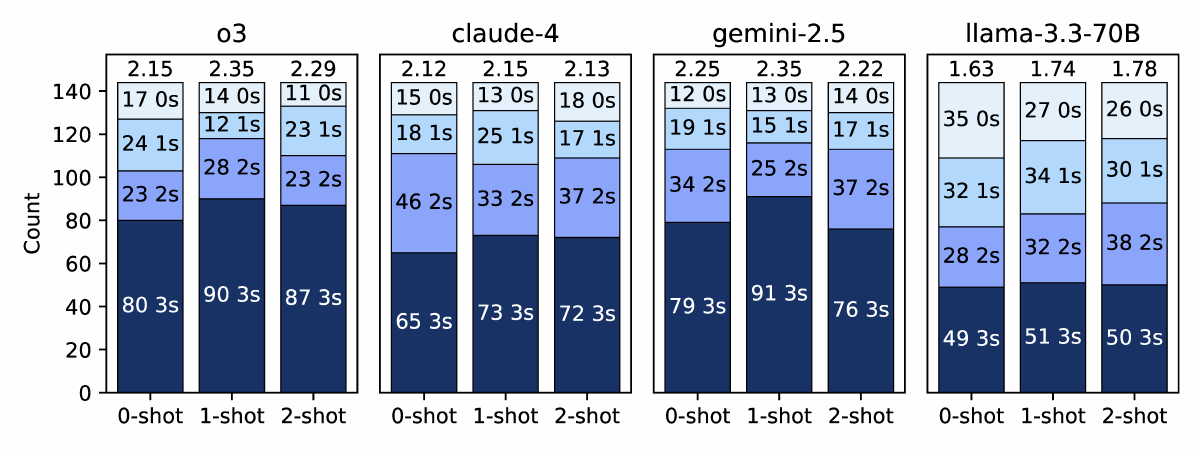}
\caption{Step-Cloze Grades with 0-, 1-, and 2-shot prompting.  Grades could be 0 (worst), 1, 2, or 3 (best). Mean grades appear over each bar. The three reasoning models outperformed the baseline llama-3.3-70B.}
\label{fig_step_cloze}
\end{figure*}

\subsection{Adversarial Step Goal-Failure Verification}

An automated approach to finding tax strategies must be able to identify strategies that do \textit{not} work.  So, we repeat the goal-verification task using each strategy's adversarial step swapped into the strategy.  These steps were initially drafted so that replacing a specified correct step with the adversarial step would keep the strategy from working (i.e. achieving all listed goals).  But the LLMs proved remarkably resistant to finding that a strategy does not work.  The LLMs would make entirely unwarranted factual assumptions to allow the strategy's goals to be met.

To address this, we engaged in an iterative process, akin to what was done with improving the analysis steps to make the models find them viable.  We improved the adversarial steps and expressly eliminated unwarranted factual assumptions to minimize the number of strategies the models found \textit{correct}, until all three reasoning models returned 0/36 as being correct. During this iterative process, no attempt was made to monitor the LLMs' performance on the strategies being \textit{viable}.

For this test, we always run without our analyses, which refer to facts in the actual strategy without the adversarial step.  So, there were only two variants of this test: viable or correct.  The results are in the two rightmost columns of Table~\ref{tab:goal_verification}. 

No domain expert would find the strategies with the adversarial steps either viable or correct, and so a model as capable as a domain expert would have 0/36 in both rightmost columns.  We see that the models all have some tendency towards finding a strategy viable.  Among the reasoning models, claude-4 found fully 20 strategies with adversarial steps viable, indicating agreement bias that might limit its usefulness.  Our baseline llama-3.3-70B model showed much stronger agreement bias than any of the reasoning models, demonstrating the benefits of size and reasoning on this task.   

\subsection{Step-Cloze Task}

An automated approach to finding tax strategies must be able to fill in a strategy's details.  So, we test LLMs' abilities to fill in partially specified strategies, so as to meet the goals.  In the 0-shot version of this task, the single prompt consists of the authorities, background, and goals, plus the strategy steps with one of the steps replaced with ‘‘[BLANK]", with the instructions ‘‘Come up with a strategy step that would replace [BLANK] and would meet the goals, given the authorities, background facts, and other steps."  (In this task, we never include the analyses, which refer to the strategy steps.)

Unlike the already-described tasks, which ask LLMs for Yes/No answers, there are varying degrees of similarity between possible strategy steps.  Thus, we use a 0-to-3 grading rubric: 3 for answers semantically identical to the blanked-out step, 2 for answers leaving out or adding some important detail, 1 for answers differing in fundamental respects, and 0 for answers with no semantic overlap.  We developed a prompt that details this grading rubric, with several examples; this prompt is on GitHub.  We had both a human domain expert and o1 use this rubric to grade a held-out split of responses from claude-3.5.  We found a Spearman's $\rho$ of 0.89 between o1 and the human grades, suggesting that this automatic grading is reliable.  We use o3 and this rubric to do the grading for all step-cloze results.

We chose two of the 36 strategies -- and one blanked-out step within each of the two -- for the 1-shot or 2-shot exemplars.  The remaining 34 strategies have a total of 144 steps.  We ran the 144 in 0-shot, 1-shot, and 2-shot forms.

Figure~\ref{fig_step_cloze} has the results, which show the LLMs have mediocre performance, at best getting 3s on slightly over half the step-clozes.  This is surprising, since all the strategies were derived from public-domain sources that were almost certainly part of the LLMs' training corpora.  We conclude that any memorization by the models of the strategies is limited, perhaps because the public sources intentionally keep the details of the strategies opaque.

Interestingly, all three reasoning models show a substantial improvement going from 0-shot to 1-shot, but then a decrease going from 1-shot to 2-shot. All three reasoning models substantially outperformed the baseline llama-3.3-70B, indicating that reasoning and size improve performance on this task.  

\subsection{From-Scratch Strategy Generation}
\label{sec_from_scratch}

The just-discussed step-cloze task asks LLMs to fill in missing steps in a partially specified tax strategy; now we discuss asking LLMs to generate entire tax strategies from scratch, with zero steps given or even suggested in the prompt.  Obviously, simply prompting an LLM with ‘‘Create a tax strategy" is not useful, since even the most creative human tax advisors need to understand their clients' goals and background, and which tax authorities might apply to them.

So, we use the authorities, background, and goals from our 36 strategies as seeds to give the LLMs a starting point for generating strategies.  Our prompt says ‘‘You will be coming up with a tax strategy that meets specified goals, given background facts and particular tax-law authorities that the strategy should employ to reach the goals," and then gives only the authorities, background, and goals.

Evaluating the LLMs' free-form answers is exceptionally expensive, since it requires a tax-law domain expert to do time-consuming legal research.  When an LLM generates the same strategy as we have in the dataset, the LLM has clearly succeeded.  But when the LLM generates a somewhat or wholly different strategy, as typically happened, determining whether it meets the goals given the authorities and background requires domain expertise and legal research.

At the time we ran this experiment, o1-preview and claude-3.5 were among the most advanced publicly available LLMs.  We ran against these two (specifically, claude-3-5-sonnet-20240620 and o1-preview-2024-09-12), resulting in $36 \times 2 = 72$ strategies to grade.

We used a 0-to-3 grading rubric.  A generated strategy got 3 if it works legally based on the authorities given, meets the goals, and accommodates basic economic and commercial reality.  Crucially, a strategy could get 3 \textit{even if} totally different from the strategy steps in our dataset.  Sometimes there are multiple legal strategies that meet the same goal.   A strategy got 2 if either it was a minor modification away from working, or it ignored some legal or economic issue.  A strategy got 1 if it would require substantial modifications to work, or if it would have been a 2 but made no use (implicit or explicit) of one of the authorities given.  A strategy that could not work even with substantial modifications got 0.

\begin{figure} 
\includegraphics[width=3in]{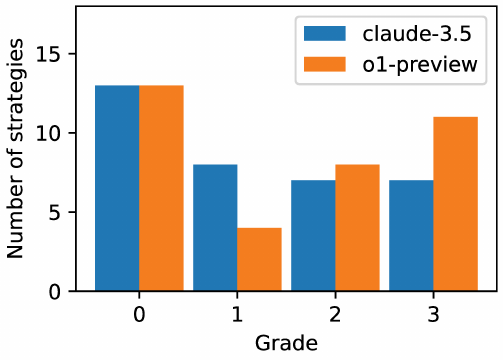}
\caption{Domain expert grades of the strategies LLMs generated from scratch. Grades could be 0 (worst), 1, 2, or 3 (best). O1-preview had more strategies than claude-3.5 with the highest grade.}
\label{fig_grading_fromscratch}
\end{figure}

Figure~\ref{fig_grading_fromscratch} contains the results.  The o1-preview strategies mostly got either a 0 or 3, whereas claude-3.5 had a somewhat more even distribution.  On average, o1-preview performed better, with a mean grade of 1.47 (versus 1.23 for claude) and with substantially more strategies getting a 3 (11 versus claude's 7).

Both models' poor performance also suggests that any memorization during LLM training of the publicly available strategies was limited.  If o1-preview or claude-3.5 had memorized a strategy, they could simply regurgitate it in response to our prompt and get a 3.  But they both got 3s on less than one-third of the strategies.

Notably, one of the strategies generated by o1-preview that got a 3 appears to be entirely novel. Figure~\ref{fig_novel} on the first page depicts this strategy.  On GitHub, in the file \texttt{Novel\_Tax\_Strategy\_17.pdf}, we include the full prompt we passed to o1-preview, as well as o1-preview's full response, plus a legal discussion of how o1-preview's strategy differs substantially from the closest previously known strategy.  Like all novel tax strategies, it uses existing legal components in unexpected ways.  The prompt that caused this response to be generated was based on strategy 17, meaning that the prompt contained the authorities, background, and goals from strategy 17, plus the instruction to come up with a strategy based on them.  Strategy 17 involves common trust funds (which are governed by tax code \S584) and financial derivatives contracts called straddles.  

\begin{figure*}[t]
\centering
\includegraphics[width=0.85\linewidth]{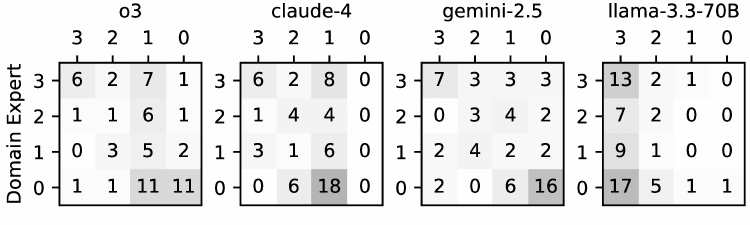}
\caption{Confusion matrices showing how well models' grading of free-form strategies compares to the domain-expert grades.}
\label{grading_fromscratch_confusion}
\end{figure*}

In o1-preview's strategy the taxpayer would become the largest participant in a common trust fund with one or more other parties.  Then, the common trust fund would enter into a straddle, which is a two-part financial instrument that ends up with one ‘‘leg" winning (i.e. going up in value) and the other ‘‘leg" losing (i.e. going down in value by an identical amount).  In step 1, the common trust fund disposes of the losing leg, passing the tax loss to the taxpayer and the other parties.  In step 2, the other parties contribute money to increase their proportionate interest in the common trust fund.  In step 3, in a later accounting period but within the same taxable year as the previous two steps, the common trust fund disposes of the winning leg, passing most of the gains to the other parties.  

By disposing of both the winning and losing legs in the same tax year, the strategy cleverly avoids the restrictions on straddles enacted by the U.S. Congress in tax code \S1092.  The taxpayer gets most of the loss from the losing leg, but little of the gain from the winning leg.  The loss reduces taxes by much more than the gain increases taxes.  

We thoroughly searched for case decisions, IRS documents, or other documents suggesting this strategy.  We found none.  While it is theoretically possible that o1-preview's training data included nonpublic tax-planning documents, that is unlikely given how tightly tax advisors and taxpayers guard their planning.  Thus, it appears o1-preview demonstrated a capacity to come up with novel tax strategies. 

\subsection{Model Grading of Strategies Generated From Scratch}

Our goal-verification task measures LLM abilities to correctly identify carefully curated viable strategies, while our adversarial-step task measures LLM abilities to filter out strategies curated to be not viable.  But what about strategies that have not been curated, such as those generated by other LLMs?

We tested all four models on their ability to grade a subset of the strategies generated by o1-preview and claude-3.5.  The prompt was the strategy plus the text of the rubric used for the grades reported in Figure~\ref{fig_grading_fromscratch}.  The confusion matrices showing how the models' grades compared with domain expert grades are in Figure~\ref{grading_fromscratch_confusion}.  

We see poor agreement of the LLM grades with the human domain expert grades.  The Spearman's $\rho$ is 0.48 for o3, 0.30 for claude-4, and 0.48 for gemini-2.5.  These are much lower than the Spearman's $\rho$ of 0.89 for step-cloze grading, likely because grading a whole strategy is harder than grading a single step.  

Claude-4 is a generous grader: it refused to give a grade of 0 to any strategies and gave a higher grade than the domain expert on fully half the strategies. As with the adversarial step goal-failure test, claude-4 shows agreement bias. And, again, we see the baseline llama-3.3-70B showing even greater agreement bias.  

\section{Dataset and Code Release}

\label{dataset_release}

All our code is on GitHub, as are five of the 36 tax strategies. Our code fully runs against these five strategies.  The five include the one that provided the seed for o1-preview's apparently novel strategy, plus the two used as exemplars for 1-shot and 2-shot step-cloze.  

We hold back the remaining 31 strategies for two reasons.  First, we want to keep the 31 from being seen by future LLMs during training, which would diminish their utility for future research.  Second, we want to keep the full dataset from researchers who aim to use AI to help taxpayers craft tax minimization strategies.  

We have shared all 36 with IRS researchers. We are willing to provide access to all 36 under a Data Use Agreement (DUA) to qualified researchers who, like us, aim to help fight tax minimization strategies. Interested parties should contact the corresponding author.

\section{Conclusion}

We tested the ability of LLMs to understand, evaluate, and generate tax-minimization strategies.  These tests produced interesting observations.  The models are responsive to different legal standards: across tasks and LLMs, questions posed using the ‘‘viable" standard got higher numbers of affirmative responses than those using the more-stringent ‘‘correct" standard.  Both o3 and claude-4 are more likely to find a strategy viable when given human-written legal analysis, while gemini-2.5 is not.  The reasoning models, and particularly claude-4, show agreement bias, towards saying strategies are viable even when the strategy is designed not to work.  On all tasks, the baseline non-reasoning model llama-3.3-70B performed worse and showed even greater agreement bias, showing the benefits of reasoning and size.  

The reasoning models are able to fill in a missing step roughly half the time and generate fully-working from-scratch strategies less than one-third of the time.  But one of these fully-working generated strategies appears to be entirely novel, suggesting great potential for LLMs in improving tax administration.

\section{Limitations}

Our dataset is small, just 36 strategies, due to the domain-expert time required to distill them from opaque public sources.  Our dataset is U.S.-centric and our findings may not generalize to other jurisdictions.  Constructing the dataset and grading the generated outputs often required legal judgment calls, which are subject to debate.  We tested only four models, although we could have tested others.

\section*{Ethics Statement}

Research towards using AI to find tax strategies can be used to help either tax agencies or taxpayers wishing to minimize their taxes.  We aim to help tax agencies, specifically the U.S. Internal Revenue Service (IRS). But our research might be used to aid tax minimization.  We attempt to ameliorate this risk by publicly releasing only five of the 36 strategies and by offering access to the remaining 31 under a data use agreement (DUA) only to other researchers interested in helping to fight tax abuse.  

The Shelter Check dataset was created entirely by applying domain expertise to publicly available sources.  Accordingly, neither its creation nor its release breach either taxpayer privacy or IRS confidentiality.  

\section*{Acknowledgments}

We would like to thank Gregory Deyesu for valuable assistance. This work has been supported by the U.S. National
Science Foundation under grant 2204926. We would like to recognize MITRE Corporation for their valuable support, assistance, and research support. Any
opinions, findings, and conclusions or recommendations expressed in this article are those of the
authors and do not necessarily reflect the views
of the National Science Foundation or MITRE.

\bibliography{aaai2026}

\end{document}